\documentclass[sigconf,natbib=true,anonymous=false,screen]{acmart}

\usepackage{graphicx}
\usepackage{amsmath}
\usepackage{tcolorbox}
\usepackage{colortbl}
\usepackage{xcolor}
\usepackage{makecell}
\usepackage{pifont}
\usepackage{algorithm}
\usepackage{algorithmic}
\usepackage{bm}
\usepackage{multicol}
\usepackage{multirow}
\usepackage[normalem]{ulem}
\useunder{\uline}{\ul}{}
\usepackage{caption}
\usepackage{enumitem}

\AtBeginDocument{%
  }

\title{DiffuRank: Effective Document Reranking with Diffusion Language Models}

\author{Qi Liu}
\email{qiliu6777@gmail.com}
\affiliation{%
  \institution{Renmin University of China}
  \city{Beijing}
  \country{China}
}

\author{Kun Ai}
\email{aikun@ruc.edu.cn}
\affiliation{%
  \institution{Renmin University of China}
  \city{Beijing}
  \country{China}
}

\author{Jiaxin Mao}
\authornote{Corresponding author.}
\email{maojiaxin@gmail.com}
\affiliation{%
  \institution{Renmin University of China}
  \city{Beijing}
  \country{China}
}

\author{Yanzhao Zhang}
\email{zhangyanzhao.zyz@alibaba-inc.com}
\affiliation{%
  \institution{Alibaba Group}
  \city{Hangzhou} 
  \country{China}
}

\author{Mingxin Li}
\email{linqi.lmx@alibaba-inc.com}
\affiliation{%
  \institution{Alibaba Group}
  \city{Hangzhou}
  \country{China}
}

\author{Dingkun Long}
\email{dingkun.ldk@alibaba-inc.com}
\affiliation{%
  \institution{Alibaba Group}
  \city{Hangzhou}
  \country{China}
}

\author{Pengjun Xie}
\email{pengjun.xpj@alibaba-inc.com}
\affiliation{%
  \institution{Alibaba Group}
  \city{Hangzhou}
  \country{China}
}

\author{Fengbin Zhu$^*$}
\email{fengbin@nus.edu.sg}
\affiliation{%
  \institution{National University of Singapore}
  \country{Singapore}
}

\author{Ji-Rong Wen}
\email{jrwen@ruc.edu.cn}
\affiliation{%
  \institution{Renmin University of China}
  \city{Beijing}
  \country{China}
}

\renewcommand{\shortauthors}{Liu, et al.}

\begin{document}

\begin{abstract}
Recent advances in large language models (LLMs) have inspired new paradigms for document reranking, in which ranking is formulated as a generation task by directly prompting LLMs to produce an ordered list of document ids. 
While this generation-based paradigm enables rerankers to better exploit the reasoning and contextual understanding capabilities of LLMs beyond traditional scoring functions, most existing LLM-based rerankers rely on autoregressive generation, which limits their efficiency and flexibility. In particular, token-by-token decoding incurs high latency and complicates output format control, while the fixed left-to-right generation order causes early prediction errors to propagate and is difficult to revise.

To address these limitations, we explore the use of diffusion language models (dLLMs) for document reranking and propose \textbf{DiffuRank}, a reranking framework built upon dLLMs. Unlike autoregressive models, dLLMs support more flexible decoding and generation processes that are not constrained to a left-to-right order, and enable parallel decoding, which may lead to improved efficiency and controllability. 
Specifically, we investigate three reranking strategies based on dLLMs: 
(1) a \emph{pointwise} approach that uses dLLMs to estimate the relevance of each query-document pair; 
(2) a \emph{logit-based listwise} approach that prompts dLLMs to jointly assess the relevance of multiple documents and derives ranking lists directly from model logits; 
and 
(3) a \emph{permutation-based listwise} approach that adapts the canonical decoding process of dLLMs to the reranking tasks. 
For each approach, we design corresponding training methods to fully exploit the advantages of dLLMs. We evaluate both zero-shot and fine-tuned reranking performance on multiple benchmark datasets. Experimental results show that dLLMs achieve performance comparable to, and in some cases exceeding, that of autoregressive LLMs with similar model sizes. These findings demonstrate the promise of diffusion-based language models as a compelling alternative to autoregressive architectures for document reranking.\footnote{The code is available at \url{https://github.com/liuqi6777/DiffusionRank}.}
\end{abstract}




\maketitle

\section{Introduction}

Document reranking plays a central role in information retrieval (IR) systems by refining an initial set of retrieved documents to produce the final ranked results shown to users.
Traditional learning-to-rank methods have been extensively studied, yet they are fundamentally constrained by predefined feature engineering and limited modeling of semantic dependencies across documents~\cite{liu2009ltr}. With the development of pre-trained language models, the cross-encoders have become the most popular and effective rerankers~\cite{reimers2019sentence,nogueira2019document,nogueira2020monot5}. Recently, the emergence of large language models (LLMs) has radically transformed this landscape~\cite{openai2024gpt4,yang2025qwen3}. Equipped with powerful contextual reasoning and generative abilities, LLMs can encode rich inter-document relationships and directly generate a ranked sequence of document identifiers, offering a new, generation-based perspective on reranking and demonstrating the state-of-the-art performance on IR benchmarks~\cite{zhu2023llm4ir, sun2023rankgpt}.

However, the majority of LLM-based ranking models are still grounded in the auto-regressive architecture of conventional LLMs, which generates document identifiers one by one, and each step is conditioned on the previously generated outputs. While well aligned with the generative capability of LLMs, this decoding paradigm introduces two key limitations. First, the strict sequential dependency prevents full parallelism, leading to substantial inference latency in large-scale reranking~\cite{reddy2024first, liu2025perank}. Second, ranking quality can degrade since errors made early in the sequence may be propagated and amplified in later steps.

Motivated by these limitations, a growing line of studies has explored non-generative approaches that directly derive rankings from LLMs without explicit sequence generation~\cite{reddy2024first, chen2024attention,liu2025e2rank}. 
For example, FIRST~\cite{reddy2024first} proposed using the logits of the first step to obtain the ranking, ICR~\citep{chen2024attention} proposed leveraging the attention scores within LLMs, and E2Rank~\citep{liu2025e2rank} proposed modeling listwise reranking as pseudo relevance feedback augmented query and use an embedding model for reranking.
These methods achieve better efficiency while maintaining effectiveness; however, they simply treat the LLM as an encoder or feature extractor, restricting the model’s generative potential for iterative refinement.

Taken together, the above observations naturally raise the following question: how can we fully leverage the generative and reasoning capabilities of large language models for document reranking, without inheriting the limitations of autoregressive decoding? To address this question, we turn to diffusion language models (dLLMs), which have recently emerged as a promising generative alternative to autoregressive LLMs~\cite{nie2025llada, ye2025dream}. Instead of producing outputs token by token, dLLMs formulate text generation as a process of iterative denoising, starting from random noise and progressively refining predictions toward the data distribution. This iterative decoding allows for parallel decoding across all token positions and provides natural flexibility for editing and correction. 

From the perspective of document reranking, dLLMs provide several advantages. 
First, dLLMs employ a bidirectional attention mechanism, which is more conducive to capturing different correlation signals between candidate documents, and can also compute context-based logits at any position, making it more flexible during inference.
Second, it inherently supports parallel decoding during inference, thus eliminating the sequential bottleneck of autoregression and effectively mitigating the accumulated errors during decoding, also bringing higher efficiency. 

Building on these insights, we introduce \textbf{DiffuRank}, a framework that systematically explores how diffusion language models can be applied to document reranking. Rather than committing to a single modeling paradigm, DiffuRank investigates multiple ways of leveraging the generative and representational properties of dLLMs under different ranking formulations. Specifically, we propose three complementary strategies: 
(1) a \emph{Pointwise} formulation that employs dLLMs as ross-encoders to independently estimate the relevance score of each query–document pair;
(2) a \emph{Logits-based Listwise} formulation that jointly evaluates multiple documents by extracting parallel logits for relevance score from multiple masked positions;
and (3) a \emph{Permutation-based Listwise} formulation that directly produces ranking sequences, and specifically, we model this as a diffusion-style sampling process or an assignment problem.
For each approach, we further design tailored training objectives and supervision strategies. Specifically, we leverage listwise distillation losses for the first two, and leverage a structure-aware masking strategy for the training of the third.
These designs explore the advantages of bidirectional contextual modeling and parallel decoding, providing empirical insights on how to fully exploit the characteristics inherent to diffusion architectures. 

We conduct comprehensive experiments based on a representative dLLM, LLaDA-1.5~\cite{zhu2025llada15}. We evaluate its both zero-shot and fine-tuned reranking performance in two benchmarks, TREC DL~\cite{craswell2020trecdl} and BEIR~\cite{thakur2021beir}, and compare with broad LLM-based baselines. 
Experimental results demonstrate that dLLM-based rerankers can achieve performance comparable to, and in several cases exceeding, that of LLM-based methods with similar model sizes, while enabling a more flexible trade-off between effectiveness and efficiency.
% Additionally, we conducted a detailed analysis, revealing how the dLLMs rank documents.

In summary, this work contributes as follows:
\begin{itemize}[labelindent=0pt, leftmargin=*]
    \item We introduce \textbf{DiffuRank} and systematically integrate diffusion language models into document reranking tasks. To the best of our knowledge, this is the first work to utilize dLLMs for document reranking.
    \item We propose three distinct reranking approaches, including pointwise, logits-based listwise, and permutation-based listwise. 
    \item We design tailored inference and training methods for adapting dLLMs to the above three approaches.
    \item We conduct extensive experiments on multiple reranking benchmarks, demonstrating that DiffuRank can match or surpass autoregressive LLMs of comparable size.
\end{itemize}

We also believe that as dLLMs continue to mature, their iterative and parallel nature may render them especially suited for structured tasks such as ranking, which demand both global coherence and flexible refinement, and our work represents a preliminary attempt in this direction.

\section{Related Work}

\subsection{Large Language Models for Reranking}

Large language models (LLMs) like GPT~\citep{openai2024gpt4} and Qwen~\citep{yang2025qwen3} have significantly advanced information retrieval, achieving state-of-the-art performance in document ranking tasks across multiple benchmarks~\citep{sun2023rankgpt, zhu2023llm4ir, chen2024tourrank}.
Existing methods generally fall into three prompting paradigms: pointwise, pairwise, and listwise. Pointwise methods evaluate each query-document pair independently, offering efficiency but lacking cross-document comparisons~\citep{liang2022holistic,sachan2022qg,zhang2023rankinggpt,liu2024demorank}. Pairwise methods compare document pairs for a given query to determine relative relevance~\citep{qin2023pairwise}. Listwise methods instead consider the entire candidate set simultaneously and generate a ranking list based on global relevance signals~\citep{sun2023rankgpt,pradeep2023rankzephyr,liu2024sliding}. Recent studies further improve listwise reranking by refining prompting strategies or the method of outputting the ranking list~\citep{reddy2024first,liu2025perank, chen2024attention, zhang2025query, liu2025e2rank}. However, these methods are limited to autoregressive LLMs.

\subsection{Diffusion Language Models}

Discrete Diffusion Models~\citep{austin2021structured, gu2022vector, nie2025llada, ou2024your} is an emerging text modeling architecture in recent years, which is suitable for the discrete nature of text. Unlike continuous diffusion models, these models operate directly on tokens through a forward corruption and reverse reconstruction process, often leveraging masking mechanisms to model text dependencies, named masked diffusion language models (MDLMs)~\citep{gu2022vector, nie2025llada}.
Recent research has shifted focus towards scaling these architectures to match the capabilities of LLMs. Pivotal development in this trajectory, such as LLaDA~\citep{nie2025llada} and Dream~\citep{ye2025dream}, demonstrates the effectiveness of scaling the dLLMs. 
% Building on this foundation, the applicability of this paradigm is rapidly expanded. For instance, LLaDA-V~\citep{you2025llada} and MMaDA~\citep{yang2025mmada} introduced multimodal capabilities for visual and general modality understanding. 
Most significantly, LLaDA-1.5~\citep{zhu2025llada15} incorporated advanced post-training to enhance its capabilities. These advances indicate that dLLMs can tackle more complex downstream tasks. Within information retrieval, recent studies have begun to explore dLLMs for recommendation~\citep{shi2025lladarec} and generative retrieval~\citep{zhao2025diffugr}. In this work, we take the first step toward applying dLLMs to document reranking, a setting that remains largely unexplored.

\section{Preliminaries}

Different from auto-regressive LLMs, masked diffusion language models formulate text generation as a denoising process that gradually recovers a clean text from its corrupted version~\cite{austin2021structured, shi2024simplified, nie2025llada}.

\paragraph{Forward Process} Given a clean text sequence $x_0 = (x_0^1, \ldots, x_0^L)$, MDLMs define a forward noising process that progressively masks a subset of tokens. Specifically, at a chosen noise level $t \in [0, 1]$, each token is independently replaced with a mask token $\mathtt{[M]}$:
\begin{equation*}
    q(x_t \mid x_0) = \prod_{i=1}^{L} q(x_t^i \mid x_0^i), \quad
% \end{equation*}
% \begin{equation*}
    q(x_t^i \mid x_0^i) = \left\{ {\begin{aligned}
        & 1-t,  && x_t^i = x_0^i,  \\
        & t,    && x_t^i = \mathtt{[M]}. 
    \end{aligned}}
    \right.
\end{equation*}
As $t$ increases, a larger fraction of tokens are masked, resulting in a more corrupted sequence.

\paragraph{Reverse Process} The goal of the model is to learn a mask predictor $p_{\theta}$ that reconstructs $x_0$ from $x_t$. 
The objective is to minimize the expected negative log-likelihood over the masked position:
\begin{equation*}
    \mathcal{L}(\theta) = -\mathbb{E}_{t, x_0, x_t } 
    \left[ \sum_{i: x_{t}^{i}=\mathtt{[M]}} \frac{1}{t}\log p_\theta(x_0^i \mid x_t)\right].
\end{equation*}
Practically, $p_\theta$ is implemented by a Transformer architecture~\cite{vaswani2017attention} that outputs a probability distribution over the vocabulary.

\begin{figure*}[t]
    \centering
    \includegraphics[width=0.92\textwidth]{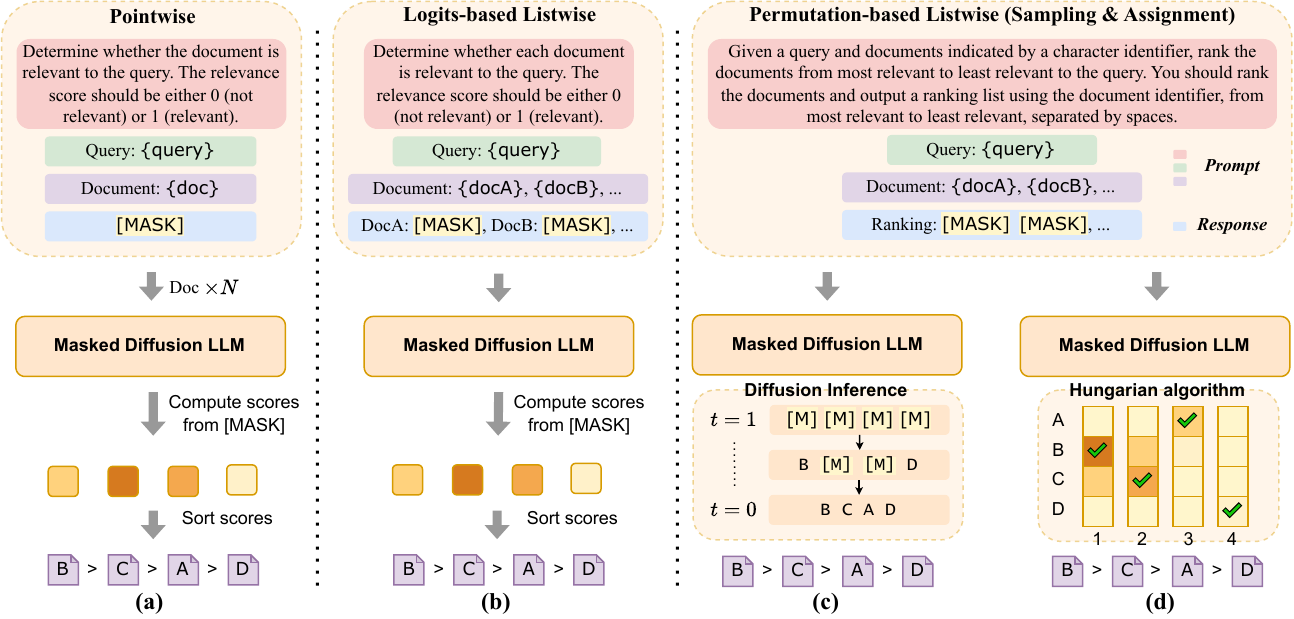}
    \vspace{-2mm}
    \caption{Overview of \textbf{DiffuRank}. \textbf{(a)} \emph{Pointwise} approach that estimates the relevance score of each query-document pair; \textbf{(b)} \emph{Logits-based Listwise} approach that jointly assesses the relevance scores of multiple documents; \textbf{(c)} and \textbf{(d)} \emph{Permutation-based Listwise} approach that directly produces the ranking list, including two inference methods: sampling and assignment.}
    \label{fig:diffusion_rank}
\end{figure*}

\paragraph{Inference}

At inference time, MDLMs approximately recover an initial fully masked sequence $\mathtt{[M]}^L$ by iteratively denoising. Specifically, in a single step from a noise level $t$ to an earlier level $s$:
\begin{equation*}
    q_{s|t}(x_s \mid x_t) = \prod_{i=1}^{L} q_{s|t}(x_s^i \mid x_t),
\end{equation*}
\begin{equation*}
    q_{s|t}(x_s^i | x_t) = \left\{ {\begin{aligned}
        & 1,              && x_t^i \neq \mathtt{[M]}, x_s^i = x_t^i,  \\
        & \tfrac{s}{t},    && x_t^i = \mathtt{[M]}, x_s^i = \mathtt{[M]} \\
        & \tfrac{t-s}{t} p_{\theta}(x^i | x_t),    && x_t^i = \mathtt{[M]}, x_s^i \neq \mathtt{[M]}. 
    \end{aligned}}
    \right.
\end{equation*}
In practice, this process is carried out via a discretized reverse procedure. At each step, the model simultaneously estimates token distributions over all masked positions, draws temporary token assignments, and then employs an adaptive remasking mechanism to select which positions should stay masked for subsequent updates. In contrast to autoregressive decoding, this paradigm resolves multiple tokens concurrently, offering greater flexibility and the potential for improved decoding efficiency.

\section{DiffuRank}

In this section, we introduce the three proposed methods that adopt dLLMs for document reranking tasks in detail, including \emph{Pointwise} approach, \emph{Logits-based Listwise} approach, and \emph{Permutation-based Listwise} approach. Figure~\ref{fig:diffusion_rank} shows an overview of these approaches.

\subsection{Pointwise Reranking}
\label{sec:pointwise}

We first explore a \emph{pointwise} formulation of document reranking using dLLMs, denoted as $\textbf{DiffuRank}_{\textit{Pointwise}}$ and shown in Figure~\ref{fig:diffusion_rank}(a). Each query–document pair is evaluated independently, and the model predicts a scalar relevance score for each pair. This formulation effectively serves as a dLLM-based cross-encoder.

\subsubsection{Inference}

Given a query $q$ and a set of $N$ candidate documents $\{d_1, \ldots, d_N\}$, we construct a prompt that asks the model to assess whether each document $d_i$ is relevant to the query. The output space is restricted to a binary relevance label, $\{0, 1\}$, corresponding to non-relevant and relevant, respectively. Formally, the input sequence is defined as:
\[
x_i = \texttt{Prompt}(q, d_i) \,\Vert\, \mathtt{[M]}, i = 1, ..., N
\]
where $\mathtt{[M]}$ is a mask token appended at the end of the sequence.  

Under the masked diffusion language model, we perform a single denoising step and obtain the logits at the masked position. Let $p_\theta(y \mid q, d_i)$ denote the predicted probability over the vocabulary at the masked token. We extract the probabilities corresponding to the tokens ``0'' and ``1'', denoted as $p_0$ and $p_1$, respectively. The final relevance score is computed as $s(q, d_i) = \frac{p_1}{p_0 + p_1}$, which yields a normalized scalar score in $[0,1]$ and can be directly used for ranking documents associated with the same query.

This inference procedure treats the diffusion language model as a bidirectional cross-encoder: the query and document attend to each other freely, and the relevance score is derived from the model’s belief at a designated masked position rather than from autoregressive generation.

\subsubsection{Training}

For training the pointwise reranker, we adopt permutation distillation following  RankGPT~\cite{sun2023rankgpt}. Given a query and a set of candidate documents, a teacher model produces a listwise ranking distribution by considering multiple permutations of the document list. The diffusion-based pointwise model is then trained to distill this supervision. Concretely, we optimize the model using either (i) a listwise Cross-Entropy loss derived from the teacher’s relevance labels:
\[
\mathcal{L}_{\text{CE}} = - \sum \mathbf{1}_{\text{rank=1}} \log \left( \frac{e^{s(q, d_i)}}{\sum_{j=1}^N e^{s(q, d_j)}}  \right),
\]
or (ii) a pairwise RankNet-style loss~\cite{burges2005ranknet} constructed from relative relevance preferences of teachers:
\[
\mathcal{L}_{\text{RankNet}} = \sum_{i=1}^N \sum_{j=1}^N \mathbf{1}_{\text{rank}_i < \text{rank}_j} \log \left( 1 + e^{s(q, d_i) - s(q, d_j)} \right),
\]
where $\text{rank}_i$ is the rank position of $i$-th document labeld by a teacher models such as RankGPT.
These objectives encourage the model to assign higher relevance scores to documents preferred by the teacher, while remaining compatible with the pointwise scoring formulation.

\subsection{Logits-based Listwise Reranking}
\label{sec:logits_listwise}

Beyond independent relevance estimation, we next consider a logits-based listwise formulation that jointly evaluates multiple candidate documents within a single forward pass, denoted as $\textbf{DiffuRank}_{\textit{Logits-List}}$. As in Figure~\ref{fig:diffusion_rank}(b), this approach leverages the bidirectional attention and parallel prediction capability of diffusion language models, allowing relevance signals to be inferred for all documents simultaneously without explicit sequence generation.

\subsubsection{Inference}

Given a query $q$ and a set of $N$ candidate documents $\{d_1, \ldots, d_N\}$, we construct a listwise prompt that assigns each document a unique numeric identifier and asks the model to determine its relevance to the query. The assistant response is structured as a list of placeholders, one per document, each replaced by a single mask token:
\[
x = \texttt{Prompt}(q, \{d_i\}_{i=1}^N) \,\Vert\, (\texttt{Doc }1{:}\ \mathtt{[M]}, \ldots, \texttt{Doc }N{:}\ \mathtt{[M]}).
\]

After a single denoising step, the diffusion language model produces logits at all masked positions in parallel. Similar to the Pointwise method, for the $i$-th document, we extract the probabilities of the binary relevance tokens ``0'' and ``1'', denoted as $p_0^{(i)}$ and $p_1^{(i)}$. The relevance score for document $d_i$ is then computed as:
\[
s(q, d_i) = \frac{p_1^{(i)}}{p_0^{(i)} + p_1^{(i)}}.
\]
Documents are ranked in descending order of $s(q, d_i)$. Since all scores are obtained from a single forward, the inference procedure is fully parallel and avoids the sequential dependency inherent in autoregressive decoding.

In practice, since current dLLMs are not optimized for long context, to handle long candidate lists that exceed the context window, we adopt the sliding window strategy introduced in RankGPT~\cite{sun2023rankgpt}. Specifically, documents are processed in overlapping windows, and local rankings within each window are iteratively merged to produce a complete global ranking list.

\subsubsection{Training}

The logits-based listwise model is also trained using permutation distillation with listwise Cross-Entropy or RankNet loss. Both losses are computed directly from the relevance scores derived at the masked positions, ensuring that training remains consistent with the inference procedure. By jointly predicting relevance logits for multiple documents in parallel, this formulation more effectively exploits the bidirectional contextual modeling of different documents, while retaining high inference efficiency.

\subsection{Permutation-based Listwise Reranking}
\label{sec:perm_listwise}

We further investigate a permutation-based listwise reranking approach that directly produces a permutation of document identifiers, analogous in spirit to RankGPT~\cite{sun2023rankgpt}, but adapted to the dLLMs. Unlike autoregressive listwise rerankers that decode identifiers from left to right, our method performs multi-token prediction in parallel, enabling controllable output refining and processing.

\subsubsection{Inference}
\label{sec:perm_infer}

Given a query $q$ and a set of $N$ candidate documents $\{d_1,\ldots,d_N\}$, we assign each document a unique identifier from a fixed alphabet (e.g., ``A'', ``B'', \ldots). The prompt asks the model to output a ranking sequence of identifiers from most to least relevant. 
We reserve a fully masked response $r_{1} = \mathtt{[M]}^{N}$ with $N$ slots for ranking and reconstruct $r_0$ as the ranking list through the reverse process.
Since dLLMs naturally perform multiple token predictions, we design different strategies to obtain the permutation.

\paragraph{Sampling the permutation}

A direct method to obtain the permutation is generating it through the reverse process, i.e., a sampling algorithm, of dLLMs. The vanilla reverse process used by dLLMs like LLaDA~\cite{nie2025llada} is typically a low-confidence remasking strategy. At each timestep, the model greedily predicts tokens for masked positions and assigns a confidence score to each position, defined as the probability of the selected token. Then, a subset of tokens with the lowest confidence is remasked to allow further refinement in later steps.\footnote{For more details, please refer to Algorithm 5 of the original LLaDA paper~\cite{nie2025llada}.}

However, directly applying it to permutation-based ranking may yield invalid outputs (e.g., duplicates or misses). Furthermore, it even allows duplicates \emph{within the same timestep} if multiple masked positions independently choose the same identifier. 
Therefore, we propose explicitly avoiding this by \emph{dynamically updating} the feasible identifier set and adopting a \emph{greedy strategy} within the step.

Concretely, at timestep $t$, let $\mathcal{M}$ be the set of masked positions in $r_t$, and let $\mathcal{U}_t \subseteq \mathcal{V}_{\text{id}}$ be the set of identifiers that have not appeared in the current partially decoded sequence $r_t$.\footnote{In practice, we treat an identifier as used if it appears in any unmasked position of $r_t$.}  $\mathcal{V}_{\text{id}}=\{\texttt{ID}_1,\ldots,\texttt{ID}_N\}$ denotes the identifier set for all candidate documents.
With the model forward, we can compute the probabilities under $\mathcal{U}_t$ for each masked position, forming a score matrix
% \[
% \bm{S}\in\mathbb{R}^{|\mathcal{M}|\times |\mathcal{U}_t|},\qquad 
% \bm{S}_{a,b} = p_\theta\!\left(\mathcal{U}_t^{(b)} \mid \rho_0, r_t, i=\mathcal{M}^{(a)}\right).
% \]
\[
\bm{P}\in\mathbb{R}^{|\mathcal{M}|\times |\mathcal{U}_t|},\qquad 
\bm{P}_{i,j} = p_\theta\!\left(r_0^i \mid \rho_0, r_t\right)_{\texttt{ID}_j},
\]
where $i \in \mathcal{M}$ and $\texttt{ID}_j \in \mathcal{U}_t$, and generally we have $|\mathcal{M}| = |\mathcal{U}_t|$.
We then create a list of all candidate triples $(\bm{P}_{i,j}, i, j)$, sort them by score in descending order, and greedily accept pairs subject to two constraints: (i) each masked position can be filled at most once, and (ii) each identifier can be used at most once in the current step. This produces a duplicate-free set of assignments that is globally prioritized by confidence. The filled tokens are written back to obtain $r_0$, after which we apply the low-confidence remasking to allow further refinement in later steps. 
In practice, we determine a sampling step $K$, and for each step, only $\frac{N}{K}$ masked positions will be filled with docids after remasking. 
The choice of $K$ allows for an effectiveness-efficiency trade-off: a smaller $K$ can complete the ranking with fewer model calls, but the effect may be worse; and vice versa.
Algorithm~\ref{alg:global_greedy_constrained} summarizes this stepwise decoding, and we denote this method as $\textbf{DiffuRank}_{\textit{Perm-Samp}}$, shown in Figure~\ref{fig:diffusion_rank}(c).

\begin{algorithm}[t]
\small
\caption{Constrained Sampling Strategy}
\label{alg:global_greedy_constrained}
\begin{algorithmic}[1]
\REQUIRE $p_\theta$, $\rho_0 = \texttt{Prompt}(q, \{d_i\}_{i=1}^N)$, $\mathcal{V}_{\text{id}}$, and sampling step $K$
\STATE Initialize $r_{1} \leftarrow \mathtt{[M]}^{N}$
\FOR{$t \leftarrow 1$ \textbf{down to} $1/K$ \textbf{step} $1/K$}
    \STATE $s \leftarrow t - 1/K$
    \STATE $\mathcal{M} \leftarrow \{i \in [N] : r_t^i = \mathtt{[M]}\}$ \\
    \IF{$|\mathcal{M}|=0$} \STATE \textbf{break} \ENDIF
    \STATE $r_0 \leftarrow r_t$
    \STATE $\mathcal{U}_t \leftarrow \mathcal{V}_{\text{id}} \setminus \{r_t^i : r_t^i \neq \mathtt{[M]}\}$ \hfill {\small // update constrained token set}
    \STATE Compute $\bm{P}\in\mathbb{R}^{|\mathcal{M}|\times |\mathcal{U}_t|}$ \hfill {\small // restrit score to the token set}
    \STATE Build $\mathcal{P} \leftarrow \{(\bm{P}_{a,b}, a, b)\}$ and sort $\mathcal{P}$ by score descending
    \STATE Initialize $\mathcal{A}\leftarrow \emptyset$ (assigned positions), $\mathcal{B}\leftarrow \emptyset$ (used identifiers)
    \FOR{each $(v,a,b)$ in sorted $\mathcal{P}$}
        \IF{$a \in \mathcal{A}$ \OR $b \in \mathcal{B}$} \STATE \textbf{continue} \ENDIF
        \STATE $r_0^{\mathcal{M}^{(a)}} \leftarrow \mathcal{U}_t^{(b)}$; \quad $c^{\mathcal{M}^{(a)}} \leftarrow v$
        \STATE $\mathcal{A}\leftarrow \mathcal{A}\cup\{a\}$; \quad $\mathcal{B}\leftarrow \mathcal{B}\cup\{b\}$
        % \IF{$|\mathcal{A}| = \min(|\mathcal{M}|,|\mathcal{U}_t|)$} \STATE \textbf{break} \ENDIF
    \ENDFOR
    \FOR{each $i$ with $r_t^i \neq \mathtt{[M]}$}
        \STATE $c^{i} \leftarrow 1$
    \ENDFOR
    \STATE $n_{\text{un}} \leftarrow \lfloor N \times (1-s)\rfloor$ \hfill {\small // number of unmasked tokens at timestep $s$}
    \FOR{$i \leftarrow 1$ to $N$}
        \IF{$c^i \in \textsc{Lowest}_{\,N-n_{\text{un}}}(\{c^j\}_{j=1}^{N})$}
            \STATE $r_0^i \leftarrow \mathtt{[M]}$ \hfill {\small // remask lowest-confidence positions}
        \ENDIF
    \ENDFOR
    \STATE $r_s \leftarrow r_0$; $r_t \leftarrow r_s$
\ENDFOR
\STATE \textbf{return} $r_0$
\end{algorithmic}
\end{algorithm}
\vspace{-2mm}

\paragraph{Permutation as Assignment Problem}

Different from the diffusion-style sampling strategy above, this variant does \emph{not} rely on iterative sampling steps. Instead, we can perform a single forward pass to obtain a global permutation in one shot by formulating decoding as an assignment problem.

Concretely, given the prompt $\rho_0$ and a ranking sequence initialized as fully masked
$r = (\mathtt{[M]}, \ldots, \mathtt{[M]})$,
we run the model once to obtain logits for all $N$ masked positions. For each rank position $i$ and document identifier $j$, we compute the probability
\[
\bm{P}_{i,j} = p_\theta\!\left(r_0^i \mid \rho_0, r_t\right)_{\texttt{ID}_j}
\]
and construct a cost matrix $\bm{C}\in\mathbb{R}^{N\times N}$ with
\[
\bm{C}_{ij} = -\log \bm{P}_{ij}.
\]
We then solve a minimum-cost bipartite matching problem:
\[
\min_{\pi \in \mathcal{S}_N} \sum_{i=1}^{N} \bm{C}_{i,\pi(i)},
\]
where $\mathcal{S}_N$ denotes the set of all permutations of $N$ elements. This optimization can be efficiently solved using the \textbf{Hungarian algorithm}~\cite{kuhn1955hungarian}, which is a combinatorial optimization algorithm that solves the assignment problem in polynomial time. In this scenario, the algorithm yields a permutation $\pi$ that assigns each document identifier to a unique rank position, and the final ranking is directly read from this assignment. We denote this method as $\textbf{DiffuRank}_{\textit{Perm-Assign}}$ and show in Figure~\ref{fig:diffusion_rank}(d).

This formulation reveals an alternative interpretation of reranking with dLLMs: the model effectively performs parallel multi-token prediction, where each rank position predicts a distribution over document identifiers, and the resulting $N\times N$ probability (or cost) matrix can be viewed as a structured prediction over permutations. The Hungarian algorithm provides one principled and simple strategy under the constraint of one-to-one assignment. Moreover, this perspective also suggests a broader design space. For example, one could incorporate additional constraints, apply alternative global objectives, or design differentiable relaxations for end-to-end optimization. In this work, we adopt the Hungarian algorithm as a straightforward instantiation to demonstrate feasibility, leaving richer strategies as an interesting direction for future research.

\subsubsection{Training}
\label{sec:perm_train}

\paragraph{SFT objective for dLLMs.}
We train the permutation-based reranker using supervised fine-tuning (SFT) under the MDLM denoising objective. Each training instance consists of a prompt (query and documents) and a target response string that encodes the ground-truth ranking identifiers. Let $x$ denote the concatenation of prompt and response, and let $\ell$ be the prompt length (so positions $i\ge \ell$ belong to the response). Following MDLM training, we sample a corruption level and mask a subset of response tokens to obtain a noisy sequence $x_t$, and optimize:
\[
\mathcal{L}(\theta) = -\mathbb{E}_{t}\!\left[\sum_{i: x_t^i=\mathtt{[M]}} \frac{1}{p_i}\log p_\theta(x^i \mid x_t)\right],
\]
where $p_i$ is the masking probability assigned to token position $i$ (used for reweighting, consistent with LLaDA-style training). In implementation, we normalize the loss by the response length to stabilize optimization across variable-length inputs.

\paragraph{Masking strategy.}

The vanilla forward process masks each response token independently with a per-sample mask probability. Concretely, we sample $t\sim \mathcal{U}(0,1)$ and set $p_{\text{mask}}=(1-\varepsilon)t+\varepsilon$, then for each response position we mask it with probability $p_{\text{mask}}$. This produces standard denoising training signals but does not exploit the structure of ranking outputs.

To better align training noise with the nature of ranking generation, we propose a fine-grained corruption process that perturbs \emph{only document identifier tokens} in the response. Specifically, we randomly mask a subset of identifiers, forcing the model to recover missing identifiers given the remaining partial ranking. The mask probability is the same as the vanilla mask strategy.

\section{Experiments}

In this section, we conduct extensive experiments to evaluate DiffuRank from multiple perspectives. Specifically, we study the following research questions:
\begin{itemize}[labelindent=0pt, leftmargin=*]
\item \textbf{RQ1:} How does the reranking performance of dLLMs compare to that of LLMs in zero-shot scenarios and training settings?
\item \textbf{RQ2:} Which reranking approach is more effective in leveraging dLLMs for document reranking?
\item \textbf{RQ3:} How do different training and inference strategies affect the reranking performance of dLLMs?
\item \textbf{RQ4:} What is the behavior of dLLMs in reranking, and what are the differences compared to LLMs?
\end{itemize}

\subsection{Experimental Setup}

\subsubsection{Benchmarks}
\label{sec:benchmarks}

We evaluate DiffuRank on two reranking benchmarks that reflect complementary retrieval scenarios.
The TREC Deep Learning (TREC DL)~\cite{craswell2020trecdl} is a standard testbed for neural reranking, featuring queries from real-world web search logs and large-scale passage and document collections. We report results on the official TREC DL 19 and 20 test sets.
BEIR~\cite{thakur2021beir} is a heterogeneous benchmark suite that covers a broad range of information retrieval tasks. Following previous work~\cite{sun2023rankgpt}, we conduct evaluations on 8 datasets that contain a relatively small number of queries, including TREC Covid, NFCorpus, Touch2020, DBPedia, SciFact, Signal1M, TREC News, and Robust04. For all benchmarks, we rerank the top-100 candidate documents retrieved by BM25~\cite{robertson2009bm25} and use NDCG@10 as the evaluation metric.

\subsubsection{Baselines}
\label{sec:baselines}

We compare DiffuRank with a broad set of strong baselines, covering both zero-shot and fine-tuned settings.

\paragraph{Zero-shot baselines.}
We first consider zero-shot rerankers based on large language models of comparable parameter scales, including Qwen3-8B~\cite{yang2025qwen3} and LLaMA3.1-8B~\cite{grattafiori2024llama3}. 
On top of these backbones, we include several zero-shot reranking paradigms, including Pointwise~\cite{liang2022holistic} and Listwise~\cite{sun2023rankgpt}. These baselines cover the dominant zero-shot reranking strategies proposed for LLMs. We also include vanilla LLaDA-1.5~\cite{zhu2025llada15} and use the same listwise method as a baseline for comparison with DiffuRank.

\paragraph{Fine-tuned baselines.}
In the supervised setting, we compare with both traditional neural rerankers and fine-tuned LLM-based methods. We include a strong \emph{cross-encoder} baseline trained with relevance supervision, including monoBERT~\cite{nogueira2019monobert} and monoT5~\cite{nogueira2020monot5}. We also consider a previous fine-tuned LLM-based listwise rerankers RankZephyr~\cite{zhuang2023rankt5}. In addition, we also fine-tune the aforementioned backbones using the same training data and training loss (i.e., listwise distillation or SFT) as DiffuRank. These baselines enable a fair comparison under matched supervision and model capacity, allowing us to assess whether diffusion-based rerankers can match or surpass autoregressive LLMs when both are trained specifically for reranking.

\subsubsection{Implementation Details}

\paragraph{Backbone Model.}
All diffusion-based rerankers are built upon LLaDA-1.5~\cite{zhu2025llada15}, which is a powerful dLLM with 8B parameters that has undergone post-training and is suitable for various downstream tasks. For longer input sequences in listwise settings, we follow LongLLaDA~\cite{liu2025longllada} and adopt RoPE~\cite{su2024roformer} scaling with a factor of 14.0 to support a context length of 16k.

\paragraph{Training Dataset and Parameters.}
We used the dataset provided by RankZephyr~\cite{pradeep2023rankzephyr}, which contains approximately 40K samples from MS MARCO~\citep{bajaj2016msmarco}, and each sample consists of a query and a maximum of 20 ranked candidate documents annotated by GPT-4. The LLM-based baselines are trained with the same dataset.
During the training, we apply LoRA adaptation with a rank of 16, an alpha of 32, and a dropout rate of 0. We train the models for three epochs using the AdamW optimizer with a constant learning rate of $1\times10^{-4}$ and weight decay of 0.01. We use gradient accumulation to achieve a total batch size of 64 under 8 NVIDIA A100 80G GPUs. We also integrate mixed precision and DeepSpeed ZeRO-3  for reducing GPU memory and acceleration.

\paragraph{Evaluation Details.}
For all LLM-based baseline methods, we use the \textsc{LLM4Ranking} framework~\cite{liu2025llm4ranking} and follow its default prompt templates and inference configurations to ensure reproducible comparisons. The listwise methods employ a sliding window strategy~\cite{sun2023rankgpt} with a window size of 20 and a step size of 10. 

\begin{table*}[h]
\centering
\small
\caption{Zeor-shot performance comparison on the BEIR benchmark. All the models rerank the top 100 documents retrieved by BM25. For $\textbf{DiffuRank}_\textit{Perm-Samp}$, we set sampling step $K=2$. We present the ranking effectiveness in terms of NDCG@10, the best result of each benchmark is \textbf{bolded}, and the second best is underlined. Superscripts denote statistically significant improvements (paired Student’s t-test with $p \leq 0.05$).}
\label{tab:zero-shot}
\begin{tabular}{@{}ll|llllllll|c@{}}
\toprule
  & \textbf{Models} & \textbf{Covid} & \textbf{NFCorpus} & \textbf{Touche} & \textbf{DBPedia} & \textbf{SciFact} & \textbf{Signal} & \textbf{News}  & \textbf{Robust} & \textbf{Avg.} \\ \midrule
\textit{a} & \textbf{BM25} & 59.47 & 33.75 & \textbf{44.22} & 31.80 & 67.89 & 33.05 & 39.52 & 40.70  & 43.43     \\ \midrule
  & \multicolumn{10}{l}{\textit{LLM-based rerankers}}  \\ \midrule
\textit{b} & $\textbf{Llama-3.1-8B}_\textit{Pointwise}$        & 78.59$^{\textit{afij}}$ & 34.93 & 29.08$^{\textit{g}}$ & 32.43 & 61.31 & 23.84 & 33.99 & \textbf{62.16}$^{\textit{acefghij}}$ & 44.54 \\
\textit{c} & $\textbf{Llama-3.1-8B}_\textit{Listwise}$         & 76.49$^{\textit{af}}$ & \underline{36.53}$^{\textit{abfh}}$ & 35.13$^{\textit{dg}}$ & 41.36$^{\textit{abdfg}}$ & 69.48$^{\textit{bg}}$ & 31.70$^{\textit{bdg}}$ & 46.19$^{\textit{abdg}}$ & 51.22$^{\textit{afi}}$ & 48.51 \\
\textit{d} & $\textbf{Qwen3-8B}_\textit{Pointwise}$            & 74.82$^{\textit{af}}$ & 35.05 & 25.47 & 32.86 & 68.28$^{\textit{bg}}$ & 25.98 & 36.85$^{\textit{b}}$ & \underline{61.57}$^{\textit{acefghij}}$ & 45.11 \\
\textit{e} & $\textbf{Qwen3-8B}_\textit{Listwise}$             & \underline{80.66}$^{\textit{acdfij}}$ & \textbf{38.28}$^{\textit{abcdfghij}}$ & 35.94$^{\textit{bdgh}}$ & \underline{42.18}$^{\textit{abdfgh}}$ & \textbf{75.91}$^{\textit{abcdfghij}}$ & 30.56$^{\textit{bdg}}$ & \textbf{48.86}$^{\textit{abdfg}}$ & 54.87$^{\textit{acfhij}}$ & \textbf{50.91} \\ \midrule
  & \multicolumn{10}{l}{\textit{dLLM-based rerankers (LLaDA-1.5-8B as the backbone model)}}   \\ \midrule
\textit{f} & $\textbf{LLaDA-1.5-8B}_\textit{Listwise}$  & 64.49$^{\textit{a}}$ & 35.09$^{\textit{a}}$ & 37.09$^{\textit{bdgh}}$ & 38.94$^{\textit{abdg}}$ & 71.26$^{\textit{abg}}$ & 32.40$^{\textit{bdg}}$ & 43.33$^{\textit{abd}}$ & 45.54$^{\textit{a}}$ & 46.02\\
\textit{g} & $\textbf{DiffuRank}_\textit{Pointwise}$       & 79.27$^{\textit{adfij}}$ & 35.63 & 23.02 & 31.60 & 61.95 & 28.04$^{\textit{b}}$ & 39.32$^{\textit{b}}$ & 53.31$^{\textit{acfhi}}$ & 44.02 \\
\textit{h} & $\textbf{DiffuRank}_\textit{Logits-List}$     & \textbf{81.08}$^{\textit{acdfij}}$ & 34.08 & 31.40$^{\textit{g}}$ & 39.79$^{\textit{abdg}}$ & 70.52$^{\textit{bg}}$ & 29.81$^{\textit{bd}}$ & \underline{48.74}$^{\textit{abdfg}}$ & 50.75$^{\textit{af}}$ & 48.27 \\
\textit{i} & $\textbf{DiffuRank}_\textit{Perm-Samp}$       & 73.08$^{\textit{af}}$ & 35.48$^{\textit{ah}}$ & 36.17$^{\textit{dgh}}$ & \textbf{43.34}$^{\textit{abcdfgh}}$ & \underline{72.90}$^{\textit{abcdg}}$ & \underline{34.17}$^{\textit{bdegh}}$ & 46.30$^{\textit{abdg}}$ & 49.16$^{\textit{af}}$ & 48.83 \\
\textit{j} & $\textbf{DiffuRank}_\textit{Perm-Assign}$     & 72.99$^{\textit{af}}$ & 36.08$^{\textit{ah}}$ & \underline{37.18}$^{\textit{bdgh}}$ & \textbf{43.34}$^{\textit{abcdfgh}}$ & 72.00$^{\textit{abg}}$ & \textbf{34.61}$^{\textit{bdefgh}}$ & 46.79$^{\textit{abdg}}$ & 51.38$^{\textit{afi}}$ & \underline{49.30} \\ \bottomrule
\end{tabular}
\end{table*}

\subsection{Main Results}

To answer \textbf{RQ1} and \textbf{RQ2}, we compare DiffuRank against various baselines on different benchmarks and different settings.

\subsubsection{Zero-shot Performance} Table~\ref{tab:zero-shot} reports the zero-shot reranking results on BEIR benchmark.

First, all rerankers substantially outperform BM25 on most of the datasets, except for Touche, confirming that both LLM-based and dLLM-based models can effectively improve retrieval quality in the zero-shot setting. According to~\cite{thakur2024systematic}, the poor performance of neural retrieval models is mainly due to the large number of short texts and unlabeled texts in the Touche dataset. 

Second, among LLM-based baselines, listwise methods are generally stronger than pointwise ones. Both Llama-3.1 and Qwen3 exhibit clear improvements when switching from pointwise to listwise prompting, with $\text{Qwen3-8B}_\textit{Listwise}$ achieving the best average performance. This highlights that explicitly modeling cross-document interactions is critical for strong zero-shot reranking.

Third, dLLM-based rerankers achieve competitive performance with strong LLM-based baselines. While $\text{DiffuRank}_\textit{Pointwise}$ is generally weaker, the listwise variants, including $\text{DiffuRank}_\textit{Logits-List}$, $\text{DiffuRank}_\textit{Perm-Samp}$, and $\text{DiffuRank}_\textit{Perm-Assign}$, consistently outperform most pointwise baselines and achieve results close to the best LLM-based reranker. Additionally, DiffuRank performs better than vanilla LLaDA-1.5, indicating the effectiveness of our proposed approaches. In particular, permutation-based methods show strong performance among all dLLM-based variants. Notably, $\text{DiffuRank}_\textit{Perm-Assign}$ achieves the highest average performance among dLLM-based methods, suggesting that explicitly modeling permutation structure as an assignment problem may provide a more stable reranking ability.

\subsubsection{Fine-tuned Performance} Table~\ref{tab:fine-tuned} presents the reranking results of fine-tuned models on TREC DL and BEIR. 

First, fine-tuning brings substantial improvements over BM25 across almost all datasets. Both LLM- and dLLM-based rerankers achieve strong performance, with listwise variants generally outperforming pointwise ones. For example, $\text{Llama-3.1-8B}_\textit{Listwise}$ improves over its pointwise counterpart on most datasets, confirming the benefit of listwise modeling.

Second, after training, dLLM-based rerankers achieve performance on par with or better than strong fine-tuned LLM baselines. Among them, the permutation-based variants are consistently the strongest. Specifically, $\text{DiffuRank}_\textit{Perm-Samp}$ and $\text{DiffuRank}_\textit{Perm-Assign}$ obtain the best average performance on BEIR, outperforming all LLM-based baselines in terms of overall average, demonstrating strong cross-domain generalization.

Third, compared with other DiffuRank variants, permutation modeling provides a clear advantage. While Pointwise and Logits-List variants already reach competitive performance, introducing explicit permutation structure further improves effectiveness. This suggests that directly learning to generate permutations offers a more suitable capability for reranking under diffusion modeling.

Overall, these results show that diffusion-based rerankers are not only competitive in the zero-shot setting but can also surpass strong fine-tuned LLM rerankers after supervised training. This validates DiffuRank as an effective and scalable alternative paradigm for neural reranking beyond conventional autoregressive architectures.

\subsubsection{Summary}

Regarding \textbf{RQ1}, across both zero-shot and fine-tuned settings, dLLM-based rerankers achieve performance that is competitive with strong LLM-based baselines.  
In the zero-shot setting, DiffuRank reaches comparable average performance to the best LLM-based baselines.  
After fine-tuning, DiffuRank further demonstrates strong effectiveness, with permutation-based variants achieving the best overall average and outperforming all evaluated LLM-based rerankers.
These results indicate that dLLMs are not only viable for reranking but can match or even surpass autoregressive LLMs when properly adapted.

\begin{table*}[]
\centering
\small
\caption{Performance comparison on the TREC DL and BEIR benchmark. All the models are fine-tuned using the MS MARCO datasets. We rerank the top 100 documents retrieved by BM25 and present the NDCG@10. For $\textbf{DiffuRank}_\textit{Perm-Samp}$, we set $K=4$ on TREC DL and $K=2$ on BEIR. The best result of each benchmark is \textbf{bolded}, and the second best is underlined. Superscripts denote statistically significant improvements (paired Student’s t-test with $p \leq 0.05$). {Some baseline results are from other papers and are provided for comparison and reference, but no significance test was performed.}}
\label{tab:fine-tuned}
\setlength{\tabcolsep}{3.5pt}
\begin{tabular}{@{}ll|ll|llllllllc@{}}
\toprule
\textit{}  & \textbf{Models} & \textbf{DL19}  & \textbf{DL20}  & \textbf{Covid} & \textbf{NFCorpus} & \textbf{Touche} & \textbf{DBPedia} & \textbf{SciFact} & \textbf{Signal} & \textbf{News}  & \textbf{Robust} & \textbf{Avg.} \\ \midrule
\textit{a} & \textbf{BM25}           & 50.58 & 47.96 & 59.47 & 33.75 & \textbf{44.22} & 31.80 & 67.89 & 33.05 & 39.52 & 40.70 & 43.43     \\ \midrule
\textit{}  & monoBERT (340M)         & 70.50 & 67.28 & 73.45 & 34.92 & 30.26 & 41.69 & 62.22 & 30.63 & 47.03 & 44.21 & 45.55     \\
\textit{}  & monoT5 (220M)           & 71.80 & 68.90 & 75.94 & 35.42 & 30.35 & 42.43 & 65.07 & 31.20 & 46.98 & 44.15 & 46.44     \\ 
\textit{}  & $\text{RankZephyr-7B}_\textit{Listwise}$          & 73.39 & 70.02 & 82.92 & 38.26    & 30.22  & 44.42   & 75.42   & 31.41  & 52.80 & 53.73  & 51.15     \\
\midrule
  & \multicolumn{12}{l}{\textit{Fine-tuned LLM-based rerankers evaluated by us}}  \\ \midrule
\textit{b} & $\textbf{Llama-3.1-8B}_\textit{Pointwise}$   & 72.55$^{\textit{af}}$ & 69.81$^{\textit{a}}$ & 83.49$^{\textit{a}}$ & 35.47 & 33.52 & 41.13$^{\textit{a}}$ & \textbf{79.59}$^{\textit{ag}}$ & 29.63 & 47.85$^{\textit{a}}$ & \underline{62.73}$^{\textit{acefgh}}$ & 51.68     \\
\textit{c} & $\textbf{Llama-3.1-8B}_\textit{Listwise}$    & \textbf{74.79}$^{\textit{af}}$ & 71.56$^{\textit{af}}$ & \textbf{87.60}$^{\textit{abef}}$ & 39.84$^{\textit{ab}}$ & 34.64 & 46.21$^{\textit{abf}}$ & 78.59$^{\textit{ag}}$ & \textbf{34.16}$^{\textit{bdg}}$ & 52.43$^{\textit{abd}}$ & 60.74$^{\textit{af}}$ & 54.28     \\
\textit{d} & $\textbf{Qwen3-8B}_\textit{Pointwise}$       & \underline{73.51}$^{\textit{af}}$ & 70.99$^{\textit{af}}$ & 85.64$^{\textit{af}}$ & \textbf{40.41}$^{\textit{abf}}$ & 35.80$^{\textit{g}}$ & 45.27$^{\textit{abf}}$ & 79.33$^{\textit{ag}}$ & 29.31 & 49.17$^{\textit{a}}$ & \textbf{66.41}$^{\textit{abcefghij}}$ & 53.92     \\
\textit{e} & $\textbf{Qwen3-8B}_\textit{Listwise}$        & 73.15$^{\textit{a}}$ & 70.75$^{\textit{af}}$ & 85.37$^{\textit{af}}$ & \underline{40.05}$^{\textit{abf}}$ & 31.73 & 45.44$^{\textit{abf}}$ & \underline{78.96}$^{\textit{ag}}$ & 32.48$^{\textit{d}}$ & 52.36$^{\textit{abd}}$ & 60.72$^{\textit{af}}$ & 53.39     \\ \midrule
  & \multicolumn{12}{l}{\textit{Fine-tuned dLLM-based rerankers (LLaDA-1.5-8B as the backbone model)}}   \\ \midrule
\textit{f} & $\textbf{LLaDA-1.5-8B}_\textit{Listwise}$  & 69.22$^{\textit{a}}$ & 65.07$^{\textit{a}}$ & 80.31$^{\textit{a}}$ & 38.67$^{\textit{ab}}$ & 40.20$^{\textit{ceg}}$ & 43.44$^{\textit{ab}}$ & 77.28$^{\textit{a}}$ & 31.93$^{\textit{d}}$ & 51.53$^{\textit{a}}$ & 56.47$^{\textit{a}}$ & 52.48     \\
\textit{g} & $\textbf{DiffuRank}_\textit{Pointwise}$  & 73.06$^{\textit{af}}$ & 69.69$^{\textit{af}}$ & 86.72$^{\textit{abf}}$ & 39.80$^{\textit{ab}}$ & 31.48 & 45.45$^{\textit{abf}}$ & 75.37$^{\textit{a}}$ & 30.24 & 52.17$^{\textit{ab}}$ & 60.23$^{\textit{af}}$ & 52.68     \\
\textit{h} & $\textbf{DiffuRank}_\textit{Logits-List}$& 72.95$^{\textit{af}}$ & 69.78$^{\textit{af}}$ & \underline{86.91}$^{\textit{af}}$ & 39.22$^{\textit{ab}}$ & 37.78$^{\textit{beg}}$ & 45.90$^{\textit{abf}}$ & 77.42$^{\textit{a}}$ & 32.62$^{\textit{d}}$ & 52.29$^{\textit{ab}}$ & 60.59$^{\textit{af}}$ & 54.09     \\
\textit{i} & $\textbf{DiffuRank}_\textit{Perm-Samp}$  & 72.92$^{\textit{af}}$ & \underline{71.72}$^{\textit{af}}$ & 86.36$^{\textit{af}}$ & 39.56$^{\textit{ab}}$ & \underline{40.74}$^{\textit{bceg}}$ & \underline{46.64}$^{\textit{abdefg}}$ & 78.27$^{\textit{ag}}$ & \underline{33.31}$^{\textit{bdg}}$ & \underline{55.20}$^{\textit{abdefgh}}$ & 60.75$^{\textit{af}}$ & \underline{55.10}     \\
\textit{j} & $\textbf{DiffuRank}_\textit{Perm-Assign}$& 73.44$^{\textit{af}}$ & \textbf{71.81}$^{\textit{af}}$ & 86.18$^{\textit{af}}$ & 39.64$^{\textit{ab}}$ & 40.54$^{\textit{bceg}}$ & \textbf{46.95}$^{\textit{abdefgh}}$ & 78.77$^{\textit{afg}}$ & 33.00$^{\textit{bd}}$ & \textbf{55.40}$^{\textit{abdefgh}}$ & 61.20$^{\textit{af}}$ & \textbf{55.21}     \\ \bottomrule
\end{tabular}
\end{table*}

For \textbf{RQ2}, among different approaches of leveraging dLLMs, permutation-based methods are consistently the most effective.  
While pointwise and logits-based listwise variants already provide competitive performance, explicitly modeling ranking as a permutation generation problem outperforms the vanilla LLaDA and yields the most stable and strongest results across both zero-shot and fine-tuned evaluations.
This suggests that incorporating permutation structure is more suitable than obtaining a relevance score for the ranking task when adapting dLLMs.

\subsection{Ablation Study}

Regarding \textbf{RQ3}, we conducted ablation studies to verify the effect of each module in DiffuRank. All experiments in this section use the fine-tuned models.

\paragraph{Influence of different training strategies}

Table~\ref{tab:ablation_training} analyzes the effect of training objectives and masking strategies.  
For both Pointwise and Logits-Listwise, RankNet consistently outperforms cross-entropy on both DL19 and DL20. The performance gap is notable (e.g., +2.5 NDCG@10 on DL19 for Pointwise), suggesting that ranking-aware supervision better aligns with reranking objectives than token-level classification. This indicates that even when the model supports flexible generation, the learning signal must explicitly encode relative preferences among documents.
For permutation-based variants, the masking strategy plays a more critical role. $\text{DiffuRank}_\textit{Perm-Assign}$ shows clear gains when using DocID Mask instead of Random Mask, especially on DL19 (73.44 vs.\ 72.06), indicating that structure-aware masking better matches the permutation prediction objective. In contrast, $\textbf{DiffuRank}_\textit{Perm-Samp}$ is less sensitive to the masking strategy, which suggests that its sampling-based formulation is more robust but also harder to benefit from a stronger structural mask during training.

\begin{table}[t]
\centering
% \small
\caption{Effect of different training losses or mask strategy on TREC DL. We report NDCG@10 as the metric.}
\label{tab:ablation_training}
\setlength{\tabcolsep}{4pt}
\begin{tabular}{@{}llcc@{}}
\toprule
 \textbf{Model} & \textbf{\makecell[l]{Training\\Strategy}} & \textbf{DL19}  & \textbf{DL20}   \\
\midrule
$\textbf{DiffuRank}_\textit{Pointwise}$   & RankNet       & 73.06 & 69.69  \\
                                              & Cross-Entorpy & 70.53 & 66.94  \\
\midrule
$\textbf{DiffuRank}_\textit{Logits-List}$      & RankNet       & 72.95 & 69.78  \\
                                              & Cross-Entorpy & 70.69 & 70.63  \\
\midrule
$\textbf{DiffuRank}_\textit{Perm-Samp}$   & DocID Mask    & 72.70 & 71.14  \\
                                              & Random Mask   & 72.55 & 71.20  \\
\midrule
$\textbf{DiffuRank}_\textit{Perm-Assign}$ & DocID Mask    & 73.44 & 71.81  \\
                                              & Random Mask   & 72.06 & 71.02  \\
\bottomrule
\end{tabular}
\end{table}

\paragraph{Influence of different inference strategies}

Table~\ref{tab:ablation} compares vanilla and constrained sampling for $\text{DiffuRank}_\textit{Perm-Samp}$. Vanilla sampling produces very low Correct\%, only around 16--17\%, meaning most generated sequences violate permutation constraints, which directly leads to poor ranking quality. In contrast, constrained sampling guarantees valid permutations, achieving 100\% Correct\%, and significantly improves performance on both datasets.
This demonstrates that enforcing structural constraints during inference is not merely a formatting issue but a key factor that determines whether the model’s generation capability can be effectively translated into reranking performance.

\begin{table}[t]
\centering
% \small
\caption{Effect of different sampling strategies used for $\text{DiffuRank}_\textit{Perm-Samp}$ on TREC DL. The sampling step $K$ is set to 4. Correct\% records the proportion of completely valid outputs (i.e., no duplicate or missing docids). {In practice, we post-process the output to ensure that the resulting permutation is valid, so even if the format is not completely correct, we can still get a ranking result.} }
\label{tab:ablation}
\setlength{\tabcolsep}{4pt}
\begin{tabular}{@{}l|cc|cc@{}}
\toprule
\multirow{2}{*}{\textbf{\makecell[l]{Sampling\\Strategy}}} & \multicolumn{2}{c|}{\textbf{DL19}}             & \multicolumn{2}{c}{\textbf{DL20}}             \\
  & \multicolumn{1}{l}{\textbf{NDCG@10}} & \textbf{Correct\%} & \multicolumn{1}{l}{\textbf{NDCG@10}} & \textbf{Correct\%} \\ \midrule
Vanilla        & 69.22 & 16.54\% & 65.07 & 17.08\%          \\
Constrained    & 72.92 & 100\%   & 71.72 & 100\%            \\ \bottomrule
\end{tabular}
\end{table}

\paragraph{Effect of sampling steps $K$}

\begin{figure}[t]
    \centering
    \includegraphics[width=\linewidth]{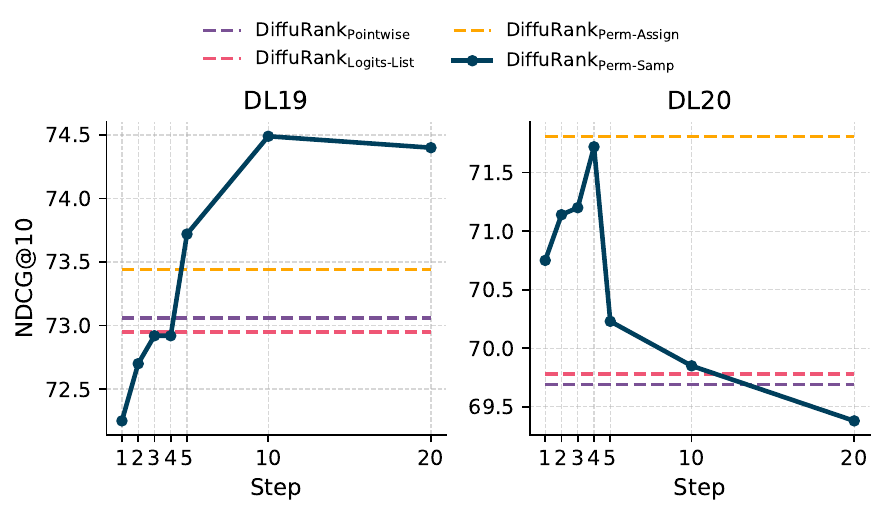}
    \caption{Effect of different sampling steps $K$ used for $\textbf{DiffuRank}_\textit{Perm-Samp}$ on TREC DL.}
    \label{fig:sampling_steps}
\end{figure}

Figure~\ref{fig:sampling_steps} further investigates the impact of sampling steps $K$ under constrained inference. On DL19, performance improves as $K$ increases from 1 to around 10, suggesting that iterative refinement can help correct noisy permutations. However, the performance slightly drops when $K$ increases to 20, indicating diminishing returns or potential over-smoothing.  
On DL20, the best performance is achieved with a small $K$ (around 3--4), and performance consistently degrades as $K$ increases. This suggests that excessive denoising steps may harm ranking quality and that the optimal number of steps is dataset-dependent. From a practical perspective, these results favor small-step inference, which is not only more efficient but also often more effective.

\paragraph{Summary}

Overall, the ablation demonstrates that dLLM-based reranking needs both training and inference design, and our proposed approaches are effective for reranking. Ranking-aware objectives and structure-aware masking are important for effective learning, while constrained inference and carefully chosen sampling steps are also essential. Simply adopting dLLMs without aligning these components leads to unstable or degraded results.

\subsection{Analysis}

To better understand how diffusion-based reranking works, we analyze the behavior of the models and attempt to answer \textbf{RQ4}. Specifically, we analyze the permutation-based methods.

\subsubsection{Analysis of Iterative Filling Dynamics}
\label{sec:analysis_dynamics}

We first investigate the \emph{iterative filling dynamics}, i.e., which ranking positions are resolved at each step, of the permutation-based generation process, revealing the behavioral pattern of the denoising process.
For each input, step $t$, and position $i$, we record whether position $i$ is \emph{first filled} at step $t$ and obtain the first-fill frequency $H(t, i)$ by aggregating over all inputs. This reflects the overall trajectory of how the ranking list is progressively constructed.
However, positions filled in earlier steps are no longer eligible in later steps. To disentangle this effect, we further compute the number of queries for which position $i$ remains masked before step $t$, denotes as $E(t, i)$. Then we can captures the conditional preference of selecting position $i$ given that it is still available: $P(t, i) = \frac{H(t, i)}{E(t, i)}$.

\begin{figure}
    \centering
    \includegraphics[width=\linewidth]{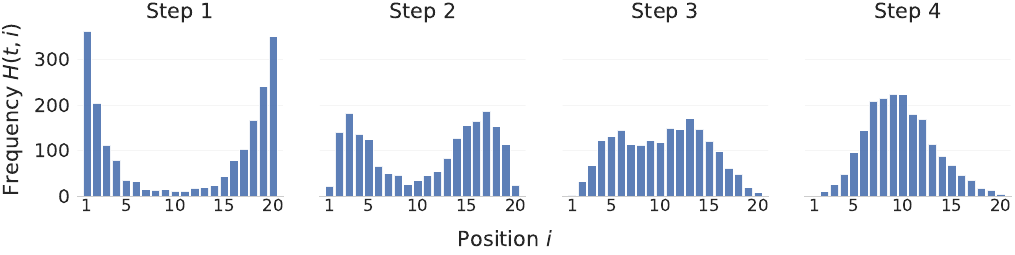}
    \includegraphics[width=\linewidth]{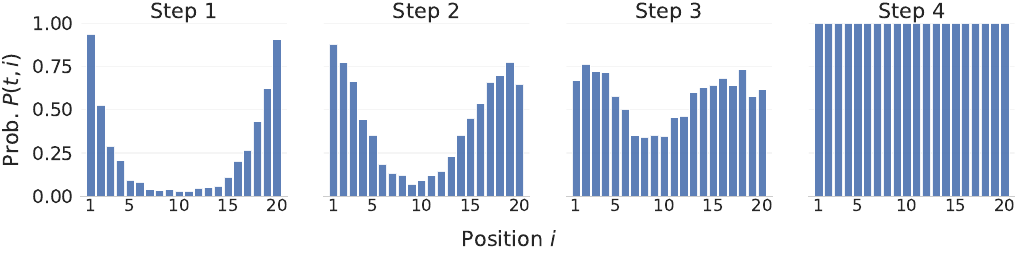}
    \caption{Iterative filling dynamics of the permutation-based reranker.}
    \label{fig:filling_dynamics}
\end{figure}

We conducted experiments on DL19 with $\text{DiffuRank}_\textit{Perm-Samp}$ and set the sampling steps to 4. 
Figure~\ref{fig:filling_dynamics} presents both the raw frequencies and the conditional probabilities across four decoding steps.
A clear and consistent pattern emerges.  
In the early steps (Step~1 and Step~2), the model exhibits a strong preference for resolving positions at both ends of the ranking list, i.e., the most relevant and the least relevant.  
In contrast, middle positions are rarely filled at this stage.  
At Step~3, the filling probability gradually shifts toward the middle region, indicating that the model focuses on refining the ordering among more ambiguous candidates. However, if there are still unfilled positions at the beginning and end, the model will still tend to fill them first.
Finally, Step~4 deterministically fills all remaining positions by design, and therefore does not carry behavioral significance, and $P(4, i)$ will constantly equal 1.
This behavior suggests a coarse-to-fine ranking strategy: the model first establishes a rough separation between highly relevant and highly irrelevant documents, and subsequently refines the ordering within the remaining uncertain region.

\subsubsection{Efficiency}

\begin{figure}[t]
    \centering
    \includegraphics[width=0.92\linewidth]{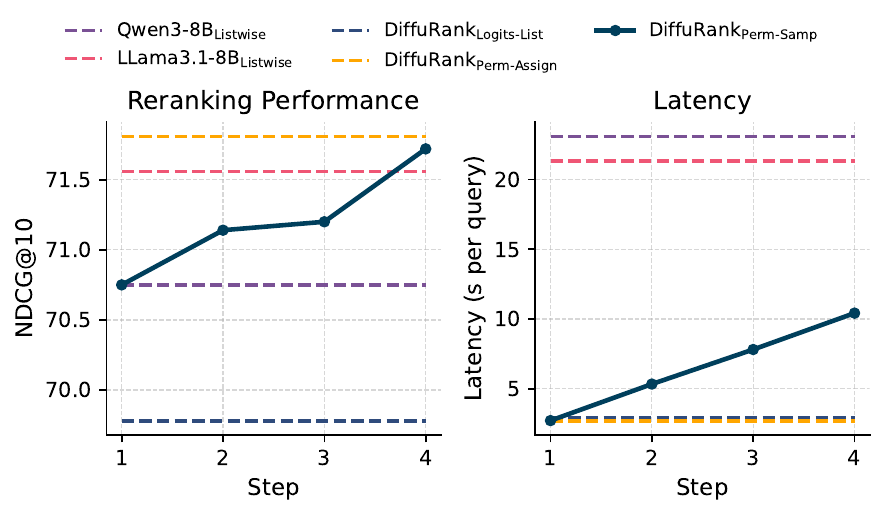}
    \caption{Efficiency analysis of listwise methods on DL20. We report the average latency of reranking top 100 documents per query, and we did not use any acceleration techniques.}
    \label{fig:latency}
    % \vspace{-2mm}
\end{figure}

Figure~\ref{fig:latency} analyzes the trade-off between effectiveness and inference latency. For a fair comparison with dLLMs, we used the vanilla Transformers implemetation for all models.
For $\text{DiffuRank}_\textit{Perm-Samp}$, increasing the sampling step improves ranking quality but incurs a clear linear increase in latency. This shows that additional denoising steps bring diminishing performance gains but non-negligible efficiency costs. Compared with strong listwise LLM baselines, permutation-based diffusion methods exhibit a more favorable efficiency–effectiveness trade-off. In particular, $\text{DiffuRank}_\textit{Perm-Samp}$ already surpasses both LLM baselines in effectiveness at $K{=}2$, while requiring less than one-third of their latency. Moreover, we can see that $\text{DiffuRank}_\textit{Perm-Assign}$ achieves strong performance with low latency, demonstrating that modeling permutation as an assignment problem can deliver both effectiveness and efficiency. These results indicate that dLLM-based reranking supports flexible trade-off, where small-step inference offers an attractive practical operating point.

\section{Conclusion}

In this work, we presented \textbf{DiffuRank}, which systematically investigates the use of diffusion language models for document reranking. By moving beyond the autoregressive generation paradigm, DiffuRank explores how the mechanisms of diffusion models can be leveraged under different ranking formulations. We studied three complementary approaches and designed corresponding inference and training strategies that align the diffusion process with the structural requirements of ranking. Extensive experiments on standard IR benchmarks demonstrate that diffusion-based rerankers can achieve performance comparable to, and in several cases exceeding, that of autoregressive LLMs with similar model sizes. 

As diffusion language models continue to mature in terms of capability, efficiency, and controllability, we believe they open new opportunities for rethinking how reranking and other problems are modeled in information retrieval.

% future work

\bibliographystyle{ACM-Reference-Format}
\bibliography{custom}

@inproceedings{liu2025perank,
  title={Leveraging passage embeddings for efficient listwise reranking with large language models},
  author={Liu, Qi and Wang, Bo and Wang, Nan and Mao, Jiaxin},
  booktitle={Proceedings of the ACM on Web Conference 2025},
  pages={4274--4283},
  year={2025}
}

@inproceedings{sun2023rankgpt,
  title={Is ChatGPT Good at Search? Investigating Large Language Models as Re-Ranking Agents},
  author={Sun, Weiwei and Yan, Lingyong and Ma, Xinyu and Wang, Shuaiqiang and Ren, Pengjie and Chen, Zhumin and Yin, Dawei and Ren, Zhaochun},
  booktitle={Proceedings of the 2023 Conference on Empirical Methods in Natural Language Processing},
  pages={14918--14937},
  year={2023}
}

@article{nogueira2019document,
  title={Document expansion by query prediction},
  author={Nogueira, Rodrigo and Yang, Wei and Lin, Jimmy and Cho, Kyunghyun},
  journal={arXiv preprint arXiv:1904.08375},
  year={2019}
}

@article{openai2024gpt4,
    title={GPT-4 Technical Report}, 
    author={OpenAI},
    journal={arXiv preprint arXiv:2303.08774},
    year={2024}
}

@article{zhu2023llm4ir,
  title={Large language models for information retrieval: A survey},
  author={Zhu, Yutao and Yuan, Huaying and Wang, Shuting and Liu, Jiongnan and Liu, Wenhan and Deng, Chenlong and Dou, Zhicheng and Wen, Ji-Rong},
  journal={arXiv preprint arXiv:2308.07107},
  year={2023}
}

@article{pradeep2023rankzephyr,
  title={RankZephyr: Effective and Robust Zero-Shot Listwise Reranking is a Breeze!},
  author={Pradeep, Ronak and Sharifymoghaddam, Sahel and Lin, Jimmy},
  journal={arXiv preprint arXiv:2312.02724},
  year={2023}
}

@article{qin2023pairwise,
  title={Large language models are effective text rankers with pairwise ranking prompting},
  author={Qin, Zhen and Jagerman, Rolf and Hui, Kai and Zhuang, Honglei and Wu, Junru and Shen, Jiaming and Liu, Tianqi and Liu, Jialu and Metzler, Donald and Wang, Xuanhui and others},
  journal={arXiv preprint arXiv:2306.17563},
  year={2023}
}

@article{thakur2021beir,
  title={Beir: A heterogenous benchmark for zero-shot evaluation of information retrieval models},
  author={Thakur, Nandan and Reimers, Nils and R{\"u}ckl{\'e}, Andreas and Srivastava, Abhishek and Gurevych, Iryna},
  journal={arXiv preprint arXiv:2104.08663},
  year={2021}
}

@article{bajaj2016msmarco,
  title={MS MARCO: A human generated machine reading comprehension dataset},
  author={Bajaj, Payal and Campos, Daniel and Craswell, Nick and Deng, Li and Gao, Jianfeng and Liu, Xiaodong and Majumder, Rangan and McNamara, Andrew and Mitra, Bhaskar and Nguyen, Tri and others},
  journal={arXiv preprint arXiv:1611.09268},
  year={2016}
}

@article{craswell2020trecdl,
  title={Overview of the TREC 2019 deep learning track},
  author={Craswell, Nick and Mitra, Bhaskar and Yilmaz, Emine and Campos, Daniel and Voorhees, Ellen M},
  journal={arXiv preprint arXiv:2003.07820},
  year={2020}
}

@article{robertson2009bm25,
  title={The probabilistic relevance framework: BM25 and beyond},
  author={Robertson, Stephen and Zaragoza, Hugo and others},
  journal={Foundations and Trends{\textregistered} in Information Retrieval},
  volume={3},
  number={4},
  pages={333--389},
  year={2009},
  publisher={Now Publishers, Inc.}
}

@article{nogueira2019monobert,
  title={Multi-stage document ranking with BERT},
  author={Nogueira, Rodrigo and Yang, Wei and Cho, Kyunghyun and Lin, Jimmy},
  journal={arXiv preprint arXiv:1910.14424},
  year={2019}
}

@article{liang2022holistic,
  title={Holistic evaluation of language models},
  author={Liang, Percy and Bommasani, Rishi and Lee, Tony and Tsipras, Dimitris and Soylu, Dilara and Yasunaga, Michihiro and Zhang, Yian and Narayanan, Deepak and Wu, Yuhuai and Kumar, Ananya and others},
  journal={arXiv preprint arXiv:2211.09110},
  year={2022}
}

@article{sachan2022qg,
  title={Improving passage retrieval with zero-shot question generation},
  author={Sachan, Devendra Singh and Lewis, Mike and Joshi, Mandar and Aghajanyan, Armen and Yih, Wen-tau and Pineau, Joelle and Zettlemoyer, Luke},
  journal={arXiv preprint arXiv:2204.07496},
  year={2022}
}

@inproceedings{zhuang2023rankt5,
  title={Rankt5: Fine-tuning t5 for text ranking with ranking losses},
  author={Zhuang, Honglei and Qin, Zhen and Jagerman, Rolf and Hui, Kai and Ma, Ji and Lu, Jing and Ni, Jianmo and Wang, Xuanhui and Bendersky, Michael},
  booktitle={Proceedings of the 46th International ACM SIGIR Conference on Research and Development in Information Retrieval},
  pages={2308--2313},
  year={2023}
}

@article{zhang2023rankinggpt,
  title={Rankinggpt: Empowering large language models in text ranking with progressive enhancement},
  author={Zhang, Longhui and Zhang, Yanzhao and Long, Dingkun and Xie, Pengjun and Zhang, Meishan and Zhang, Min},
  journal={arXiv preprint arXiv:2311.16720},
  year={2023}
}

@article{liu2024demorank,
  title={Demorank: Selecting effective demonstrations for large language models in ranking task},
  author={Liu, Wenhan and Zhu, Yutao and Dou, Zhicheng},
  journal={arXiv preprint arXiv:2406.16332},
  year={2024}
}

@article{liu2024sliding,
  title={Sliding Windows Are Not the End: Exploring Full Ranking with Long-Context Large Language Models},
  author={Liu, Wenhan and Ma, Xinyu and Zhu, Yutao and Zhao, Ziliang and Wang, Shuaiqiang and Yin, Dawei and Dou, Zhicheng},
  journal={arXiv preprint arXiv:2412.14574},
  year={2024}
}

@article{chen2024tourrank,
  title={TourRank: Utilizing Large Language Models for Documents Ranking with a Tournament-Inspired Strategy},
  author={Chen, Yiqun and Liu, Qi and Zhang, Yi and Sun, Weiwei and Shi, Daiting and Mao, Jiaxin and Yin, Dawei},
  journal={arXiv preprint arXiv:2406.11678},
  year={2024}
}

@inproceedings{reddy2024first,
  title={FIRST: Faster Improved Listwise Reranking with Single Token Decoding},
  author={Reddy, Revanth Gangi and Doo, JaeHyeok and Xu, Yifei and Sultan, Md Arafat and Swain, Deevya and Sil, Avirup and Ji, Heng},
  booktitle={Proceedings of the 2024 Conference on Empirical Methods in Natural Language Processing},
  pages={8642--8652},
  year={2024}
}

@article{zhang2025query,
  title={Query-Focused Retrieval Heads Improve Long-Context Reasoning and Re-ranking},
  author={Zhang, Wuwei and Yin, Fangcong and Yen, Howard and Chen, Danqi and Ye, Xi},
  journal={arXiv preprint arXiv:2506.09944},
  year={2025}
}

@article{chen2024attention,
  title={Attention in large language models yields efficient zero-shot re-rankers},
  author={Chen, Shijie and Guti{\'e}rrez, Bernal Jim{\'e}nez and Su, Yu},
  journal={arXiv preprint arXiv:2410.02642},
  year={2024}
}

@inproceedings{burges2005ranknet,
  title={Learning to rank using gradient descent},
  author={Burges, Chris and Shaked, Tal and Renshaw, Erin and Lazier, Ari and Deeds, Matt and Hamilton, Nicole and Hullender, Greg},
  booktitle={Proceedings of the 22nd international conference on Machine learning},
  pages={89--96},
  year={2005}
}

@article{yang2025qwen3,
  title={Qwen3 technical report},
  author={Yang, An and Li, Anfeng and Yang, Baosong and Zhang, Beichen and Hui, Binyuan and Zheng, Bo and Yu, Bowen and Gao, Chang and Huang, Chengen and Lv, Chenxu and others},
  journal={arXiv preprint arXiv:2505.09388},
  year={2025}
}

@inproceedings{nogueira2020monot5,
    title = "Document Ranking with a Pretrained Sequence-to-Sequence Model",
    author = "Nogueira, Rodrigo  and
      Jiang, Zhiying  and
      Pradeep, Ronak  and
      Lin, Jimmy",
    booktitle = "Findings of the Association for Computational Linguistics: EMNLP 2020",
    month = nov,
    year = "2020",
    pages = "708--718",
}

@article{liu2025llm4ranking,
  title={Llm4ranking: An easy-to-use framework of utilizing large language models for document reranking},
  author={Liu, Qi and Duan, Haozhe and Chen, Yiqun and Lu, Quanfeng and Sun, Weiwei and Mao, Jiaxin},
  journal={arXiv preprint arXiv:2504.07439},
  year={2025}
}

@article{nie2025llada,
  title={Large language diffusion models},
  author={Nie, Shen and Zhu, Fengqi and You, Zebin and Zhang, Xiaolu and Ou, Jingyang and Hu, Jun and Zhou, Jun and Lin, Yankai and Wen, Ji-Rong and Li, Chongxuan},
  journal={arXiv preprint arXiv:2502.09992},
  year={2025}
}

@article{zhu2025llada15,
  title={LLaDA 1.5: Variance-Reduced Preference Optimization for Large Language Diffusion Models},
  author={Zhu, Fengqi and Wang, Rongzhen and Nie, Shen and Zhang, Xiaolu and Wu, Chunwei and Hu, Jun and Zhou, Jun and Chen, Jianfei and Lin, Yankai and Wen, Ji-Rong and others},
  journal={arXiv preprint arXiv:2505.19223},
  year={2025}
}

@article{ye2025dream,
  title={Dream 7b: Diffusion large language models},
  author={Ye, Jiacheng and Xie, Zhihui and Zheng, Lin and Gao, Jiahui and Wu, Zirui and Jiang, Xin and Li, Zhenguo and Kong, Lingpeng},
  journal={arXiv preprint arXiv:2508.15487},
  year={2025}
}

@article{liu2009ltr,
  title={Learning to rank for information retrieval},
  author={Liu, Tie-Yan and others},
  journal={Foundations and Trends{\textregistered} in Information Retrieval},
  volume={3},
  number={3},
  pages={225--331},
  year={2009},
  publisher={Now Publishers, Inc.}
}

@article{ou2024your,
  title={Your absorbing discrete diffusion secretly models the conditional distributions of clean data},
  author={Ou, Jingyang and Nie, Shen and Xue, Kaiwen and Zhu, Fengqi and Sun, Jiacheng and Li, Zhenguo and Li, Chongxuan},
  journal={arXiv preprint arXiv:2406.03736},
  year={2024}
}

@article{shi2024simplified,
  title={Simplified and generalized masked diffusion for discrete data},
  author={Shi, Jiaxin and Han, Kehang and Wang, Zhe and Doucet, Arnaud and Titsias, Michalis},
  journal={Advances in neural information processing systems},
  volume={37},
  pages={103131--103167},
  year={2024}
}

@article{austin2021structured,
  title={Structured denoising diffusion models in discrete state-spaces},
  author={Austin, Jacob and Johnson, Daniel D and Ho, Jonathan and Tarlow, Daniel and Van Den Berg, Rianne},
  journal={Advances in neural information processing systems},
  volume={34},
  pages={17981--17993},
  year={2021}
}

@article{vaswani2017attention,
  title={Attention is all you need},
  author={Vaswani, Ashish and Shazeer, Noam and Parmar, Niki and Uszkoreit, Jakob and Jones, Llion and Gomez, Aidan N and Kaiser, {\L}ukasz and Polosukhin, Illia},
  journal={Advances in neural information processing systems},
  volume={30},
  year={2017}
}

@article{liu2025e2rank,
  title={E2Rank: Your Text Embedding can Also be an Effective and Efficient Listwise Reranker},
  author={Liu, Qi and Zhang, Yanzhao and Li, Mingxin and Long, Dingkun and Xie, Pengjun and Mao, Jiaxin},
  journal={arXiv preprint arXiv:2510.22733},
  year={2025}
}

@article{liu2025longllada,
  title={Longllada: Unlocking long context capabilities in diffusion llms},
  author={Liu, Xiaoran and Song, Yuerong and Liu, Zhigeng and Huang, Zengfeng and Guo, Qipeng and He, Ziwei and Qiu, Xipeng},
  journal={arXiv preprint arXiv:2506.14429},
  year={2025}
}

@article{su2024roformer,
  title={Roformer: Enhanced transformer with rotary position embedding},
  author={Su, Jianlin and Ahmed, Murtadha and Lu, Yu and Pan, Shengfeng and Bo, Wen and Liu, Yunfeng},
  journal={Neurocomputing},
  volume={568},
  pages={127063},
  year={2024},
  publisher={Elsevier}
}

@article{grattafiori2024llama3,
  title={The llama 3 herd of models},
  author={Grattafiori, Aaron and Dubey, Abhimanyu and Jauhri, Abhinav and Pandey, Abhinav and Kadian, Abhishek and Al-Dahle, Ahmad and Letman, Aiesha and Mathur, Akhil and Schelten, Alan and Vaughan, Alex and others},
  journal={arXiv preprint arXiv:2407.21783},
  year={2024}
}

@article{kuhn1955hungarian,
  title={The Hungarian method for the assignment problem},
  author={Kuhn, Harold W},
  journal={Naval research logistics quarterly},
  volume={2},
  number={1-2},
  pages={83--97},
  year={1955},
  publisher={Wiley Online Library}
}

@inproceedings{thakur2024systematic,
  title={Systematic evaluation of neural retrieval models on the Touch{\'e} 2020 argument retrieval subset of BEIR},
  author={Thakur, Nandan and Bonifacio, Luiz and Fr{\"o}be, Maik and Bondarenko, Alexander and Kamalloo, Ehsan and Potthast, Martin and Hagen, Matthias and Lin, Jimmy},
  booktitle={Proceedings of the 47th International ACM SIGIR Conference on Research and Development in Information Retrieval},
  pages={1420--1430},
  year={2024}
}

@inproceedings{gu2022vector,
  title={Vector quantized diffusion model for text-to-image synthesis},
  author={Gu, Shuyang and Chen, Dong and Bao, Jianmin and Wen, Fang and Zhang, Bo and Chen, Dongdong and Yuan, Lu and Guo, Baining},
  booktitle={Proceedings of the IEEE/CVF conference on computer vision and pattern recognition},
  pages={10696--10706},
  year={2022}
}

@article{shi2025lladarec,
  title={LLaDA-Rec: Discrete Diffusion for Parallel Semantic ID Generation in Generative Recommendation},
  author={Shi, Teng and Shen, Chenglei and Yu, Weijie and Nie, Shen and Li, Chongxuan and Zhang, Xiao and He, Ming and Han, Yan and Xu, Jun},
  journal={arXiv preprint arXiv:2511.06254},
  year={2025}
}

@article{zhao2025diffugr,
  title={DiffuGR: Generative Document Retrieval with Diffusion Language Models},
  author={Zhao, Xinpeng and Ren, Zhaochun and Zhao, Yukun and Li, Zhenyang and Zhang, Mengqi and Feng, Jun and Chen, Ran and Zhou, Ying and Chen, Zhumin and Wang, Shuaiqiang and others},
  journal={arXiv preprint arXiv:2511.08150},
  year={2025}
}

@article{reimers2019sentence,
  title={Sentence-bert: Sentence embeddings using siamese bert-networks},
  author={Reimers, Nils and Gurevych, Iryna},
  journal={arXiv preprint arXiv:1908.10084},
  year={2019}
}

\appendix

\end{document}